
\magnification=\magstep1
\baselineskip=0.48truecm
\parindent=2.5em
\parskip=0pt
\def\title#1{\centerline{\bf #1}}
\def\author#1{\bigskip\centerline{#1}}
\def\address#1{\centerline{#1}}
\noindent UCONN 94-10, December 1994
\bigskip
\title{MICROLENSING, NEWTON-EINSTEIN GRAVITY,}
\title{AND CONFORMAL GRAVITY}
\author{Philip D. Mannheim}
\address{Department of Physics, University of Connecticut,
Storrs, CT 06269-3046}
\address{mannheim@uconnvm.uconn.edu}
\smallskip
\noindent (to appear in the Proceedings of the
Fifth Annual Astrophysics Conference in Maryland)
\smallskip
We discuss some implications of the current round of galactic dark
matter searches for galactic rotation curve systematics and dynamics, and
show that these new data do not invalidate the conformal gravity program of
Mannheim and
Kazanas which has been advanced as a candidate alternative to both the standard
second order Newton-Einstein theory and the need for dark matter.
\bigskip
At the present time the search for galactic dark matter stands at an extremely
critical juncture, and while it is still far too early to draw any permanent
conclusions regarding the matter content of the Milky Way Galaxy,
it would appear that to the extent that the Galaxy actually contains dark
matter at all, not a great deal of it (and in all probability not as much of it
as had been widely surmised) may turn out to be in the form of
conventional astrophysical sources. Such standard baryonic sources essentially
fall into two categories, namely sources which are never luminous
such as brown dwarves with masses below the nuclear fusion
burning threshold or strong gravity objects such as black holes, and sources
which are in principle detectable optically but which are so faint (either by
having low absolute luminosity or by being too far away from us) so as to
simply
fall below the available level of detectability. Searches for objects in both
of these categories are now being vigorously and urgently pursued through the
gravitational microlensing observations (which can detect either category)
of the OGLE, MACHO and EROS collaborations (see e.g. the report of
B. Paczynski, these proceedings) and through the use of the recently repaired
Hubble Space Telescope (Bahcall, J. N., Flynn, C., Gould, A., \& Kirhakos, S.
1994,
ApJ. 435, L51) which now provides unprecedented optical
sensitivity in the search for faint sources. During the last year the
first results from both of these programs have been announced, with these
early results turning out to not be overly encouraging for the standard
galactic dark matter scenario. At the very minimum one can say that these
results have certainly not yet achieved their intended goal of confirming the
standard picture, while at the maximum one can say that they have actually
thrown the entire picture into doubt.

In order to identify the potential implications of these searches for
gravitational
theory, it is useful to recall some general aspects of the standard picture, a
picture
which is driven by essentially two inputs, one being the theoretical
demonstration
by Ostriker and Peebles that disk shaped Newtonian spiral galaxies are
intrinsically
unstable, and the other being the subsequent observational discovery made by
many groups
that rotational velocity curves of spiral galaxies plotted as a function of
distance $R$
from the center of each galaxy
remain tantalizingly flat far outside the optical disk region where most of the
luminous galactic matter is concentrated. Indeed, typical observed galactic
surface brightnesses behave as exp$(-R/R_0)$ with the scale length $R_0$
providing a characteristic length for a given galaxy, with rotation curves
then going out as far as 10$R_0$ in some cases. Since an exponential Newtonian
disk of matter gives a peak in the rotation curve at around 2$R_0$, beyond that
region a slow Keplerian fall off should then be expected in precisely the
kinematic region where the curves are found to be flat. Thus if the matter
distribution in fact is the same as that of the light (so that light traces
mass) then no matter what the numerical value of the mass to light ratio
$M/L$ which then normalizes the matter distribution to the surface brightness,
by
itself the luminous Newtonian contribution then fails to fit the observed shape
of the curves. (In passing we note that while the flatness of the curves
is extremely striking, the most significant
theoretical feature is the fact that the curves are actually non-Keplerian at
all). As regards the actual normalization of $M/L$ it is usually the case that
the
luminous contribution can be normalized to the Newtonian peak at 2$R_0$
so that the entire inner region of the rotation curve is then fitted purely by
the luminous disk in what are called maximum disk fits, with these fits
actually
being found to nicely accommodate the observed initial rise in the rotation
curves. One of the virtues of these maximum
disk fits is that the luminous distribution then naturally respects the
Tully-Fisher relation which is an observed correlation between the fourth power
of the peak velocity (and thus the average velocity if the rotation curve is
flat)
and the luminosity of each galaxy. However, in the maximum disk fits to
rotation
curves the $M/L$ ratios inferred from the fitting are typically found to be a
factor or so larger than that actually detected optically in the solar
neighborhood. Thus independent of what may or may not occur in the
non-Keplerian
outer region of a galaxy, we see that already in the inner disk region there is
a kind of dark matter problem for a maximal disk, since
the light does seem to be tracing the mass just as desired but with the wrong
apparent
normalization. (On the other hand if the disk is less than maximal then
something
other than the disk would be needed in the inner region to again give an inner
region dark matter problem, with it then not immediately being clear how this
extra
something would still manage to leave the Tully-Fisher
relation intact). Now in order to explain both the stability of galaxies and
the
flatness of rotation curves in the outer region the most popular approach
within the
standard Newtonian context is to invoke an ad hoc dark matter spherical halo
which is
asymptotically an isothermal sphere. (While such a halo is the most economical
option,
we note that flatness itself does not actually require a sphere - a disk with a
$1/R$
surface brightness also yields a flat rotation curve - and a less than
spheroidal halo
could still yield stability, so that the case for an explicitly spheroidal,
isothermal
halo is
actually somewhat tenuous. Moreover, since the parameters associated with the
halo
are completely unrelated to those of the disk, the halo is only able to enforce
the
Tully-Fisher relation by an artificial tuning of its parameters to those of the
disk -
the so-called conspiracy which currently lacks any explanation).
With the spherical dark halo proposal we thus encounter a second, independent
dark
matter problem, this one in the outer region, so that both the inner and outer
regions need to be explored observationally, with differing outcomes
potentially
being realizable in the two regions.

With the advent of microlensing it became possible to explicitly investigate
the above
dark matter scenario, with microlensing of the LMC in principle exploring the
dark
matter content of the halo and microlensing of the galactic bulge exploring the
dark
matter content of the disk. So far the initial searches are finding many more
lenses off the bulge than anticipated and precious few in the halo, far fewer
in fact
than widely anticipated in advance of the observations. The data thus suggest
that
the disk of the Galaxy could actually be maximal with a large $M/L$ ratio, and
that
not a great deal of matter of a conventional astrophysical nature actually
resides in the
halo (a position also supported by the failure to find copious amounts of faint
stars
in the recent HST optical survey). Thus in a sense the microlensing
observations are
nicely solving the inner region dark matter problem while simultaneously
actually making
the outer region dark matter problem more severe, since not only is the halo
dark matter not being found, but as the known amount of matter in or near the
plane
of the Galaxy is found to increase, the spherical halo would have to contain
all that
much more matter again if it is indeed going to stabilize the now much heavier
disk.
Further, these data also create a potential new problem for the fitting of
rotation
curves of dwarf galaxies, since here the standard dark matter fits favor a dark
halo
in both the inner and outer regions with the stars having to have anomalously
low $M/L$
ratios compared to spirals (low by as much as a factor of 100 in some cases).
The trend
required in dwarfs is thus completely opposite to that in regular spirals. Now
the astrophysics
which might lead to such anomalously low $M/L$ ratios for the stars in dwarfs
was never
in fact identified even before the microlensing observations (the detected
stars
themselves are not that different from those detected in spirals), and now that
we see
that large not small $M/L$ ratios are actually being preferred for normal
spiral star regions,
the dwarfs thus present a rather acute puzzle. Moreover, not only did such low
$M/L$ ratios
for the stellar disk components of dwarf galaxies not emerge from theory, they
simply
emerged from doing unconstrained fitting with an inner region dark halo giving
better
fitting than a standard, conventionally normalized inner region disk. However,
the dwarf
galaxies themselves appear to fall into two categories. There are some for
which a maximal
disk gives acceptable fitting (with a less than maximal one and a consequently
bigger halo
then just giving better fitting), and there are some for which maximal disks
completely fail in
both the shape and the normalization of the inner region rotation curves. Thus
something has to
give somewhere, and it would therefore appear to be worthwhile to again measure
the surface
brightnesses of dwarf galaxies, optimally in many filters (this would also be a
good
objective for HST), to see if some optical components have been missed or if
perhaps
some scale length values might change, so that the inner regions of
dwarf galaxies might then in fact be fitted with typical spiral $M/L$ ratios
after all.

Now given the apparent smallness of the baryonic content of the spherical halo
of the
Galaxy, there is still one option left for the standard theory, namely that the
halo be
predominantly non-baryonic being built out of some of the exotic particles now
being
theorized in elementary particle physics. (While little can currently be said
either for
or against this possibility, it would be quite remarkable if such a
non-baryonic halo
would always be able to adjust itself to the baryonic content of each galaxy
each and every
time in order to ensure the fulfillment of the Tully-Fisher relation galaxy by
galaxy).
However, unless this exotic scenario is actually realized for galaxies, we
apparently would then
have nowhere left to turn but to say that there may not actually be much of a
halo,
and that rather than being missing, the missing mass may not in fact be there
at all. Such
a step would then oblige us to have to think the unthinkable and question the
one
remaining ingredient of the standard picture, namely the validity of the
Newton-Einstein
theory of gravity itself on galactic distance scales, with the problem then
lying not with any
lack of knowledge of the astrophysical makeup of galaxies but rather with the
laws of gravity
themselves.

To look beyond Newton-Einstein (as well as to see how it is even in principle
possible to go
beyond Einstein while retaining its tested features) constitutes the program
entered into a few
years ago by Mannheim and Kazanas with their development and promotion of
fourth order conformal
gravity (for a review of this program as well as of some other candidate
alternatives to the
conventional wisdom see the report of D. Kazanas, these proceedings). One of
the main findings
of their work is the recognition that the standard second order Poisson
gravitational equation
(and accordingly its second order covariant Einstein generalization) is only
sufficient to give
Newton's Law of Gravity but not in fact necessary, a result they demonstrated
explicitly (Mannheim,
P. D., and Kazanas, D. 1994, GRG 26, 337) simply by obtaining Newton's Law in
another covariant
theory. Specifically, they found the potential $V(r)=-\beta/r+\gamma r/2$ to be
the solution to the
fourth order Poisson equation which explicitly emerges in the conformal fourth
order covariant
theory,
to thereby, in principle at least, divorce the Newtonian potential from the
standard second order
theory. Thus at the present time Einstein is only sufficient to yield Newton
but not necessary,
and the theoretical challenge that Mannheim and Kazanas threw out to the
community was to
either find some underlying fundamental principle which would make the standard
theory necessary
too (and thus actually force us to second order gravity), or to recognize
that it is the very absence of any such principle (rather than the
phenomenological issue of dark
matter) which actually opens the door to candidate alternatives even at this
late date.

Since the conformal gravity potential reduces to Newton on short enough
distance scales and then
first deviates from it galactically (for an appropriately chosen $\gamma$), the
viewpoint espoused
by Mannheim and Kazanas is that from the study of the solar system we only
measure the first few
terms in a perturbation series and that at larger distances the series may
simply differ from that
inferred from Newton-Einstein, i.e. that it may depart from the standard model
in precisely the
kinematic region where the conventional wisdom is currently having problems,
with the apparent need
for dark matter then simply stemming from having guessed the wrong series. To
explore the
conformal option fits have been made to a representative set of galactic
rotation curves using the
$V(r)=-\beta/r+\gamma r/2$ potential in conjunction with the luminous matter
distribution alone
(to thus yield a fully specified and hence falsifiable theory), and the fits
are found
(Mannheim, P. D. 1993, ApJ. 419, 150) to rival in quality standard dark matter
fits. (The fits
achieve flatness in the explored 2$R_0$ to 10$R_0$ region by an interplay
between the falling
Newtonian contribution and a rising linear contribution with the linear
contribution
mirroring that of a dark halo in this region. The fits then differ from the
standard dark matter
fits in the yet to be explored region beyond 10$R_0$ with the conformal theory
predicting a
testable eventual rise in the rotation curves). Additionally, unlike the dark
matter fits where the
two components (luminous and dark) have independent normalizations, in the
conformal theory the
linear and Newtonian potential contributions must both be integrated over the
same matter
distribution and thus must both have the same normalization. Consequently, in
order to avoid
falling below the curve at all observed distances the disk must be noneother
than
maximal. Further, in the fits, the coefficient $\gamma$ is found to be
universal for the regular
spirals considered, and intriguingly to take a value of order the inverse
Hubble radius. This
universality for galactic $\gamma$ turns out to be equivalent to the
Tully-Fisher relation
(precisely because the linear and Newtonian contributions are both
simultaneously normalized to the
one luminosity). Moreover, despite having no halo,
linear potential disks also turn out to be stable (Christodoulou, D. M. 1991,
ApJ. 372, 471). The
conformal linear potential theory thus appears capable of reproducing all the
desirable aspects of
galactic dark matter without needing the dark matter itself, and also appears
to have survived the
microlensing observations unscathed. Thus
it must currently be regarded as viable. This work has been supported in part
by the Department of
Energy under grant No. DE-FG02-92ER40716.00.

\end